\documentclass[a4paper,11pt]{article}
\pdfoutput=1 % if your are submitting a pdflatex (i.e. if you have
\usepackage{jinstpub} % for details on the use of the package, please
\usepackage{ulem}
\usepackage{threeparttable}
\title{\boldmath Development of the firmware logic validation system using the FPGA accelerator}

\author{Ryugo Mizuhiki,}
\author{Junpei Maeda,}
\author{Seiya Marumoto}

\affiliation{Department of Physics Graduate School of Science, Kobe University,\\Kobe, Hyogo, 657-8501, Japan}

\emailAdd{mizuhiki@stu.kobe-u.ac.jp}

\abstract{
Validating FPGA firmware logic used in particle physics is becoming increasingly difficult as the implementation logic scales and becomes more complex with the expansion of FPGA resources.
In order to address this issue efficiently, we have developed a firmware validation system utilizing an FPGA embedded on the PCI-express board, referred to as FPGA accelerators, produced by FPGA vendors. We developed a system that controls communication with DDR memory and runs logic on the FPGA. 
This system supports the use of firmware written in Hardware Description Language as well as High-Level Synthesis firmware, enabling rapid validation.
This paper details the developed system and its significant improvements in validation speed and flexibility compared to traditional methods. The source code is available at GitLab\footnote{\url{https://gitlab.cern.ch/rmizuhik/trigger-verification-by-accelerator}}.
}


\proceeding{Topical Workshop on Electronics for Particle Physics - TWEPP2024\\
  30 September - 04 October, 2024\\
  GLASGOW, UNITED KINGDOM}

\begin{document}
\maketitle
\flushbottom

\section{Introduction}
\label{sec:intro}

Starting in 2030, the LHC will increase the instantaneous luminosity to $5-7.5\times 10^{34} \ \rm{cm^{-2}} \rm{s^{-1}}$ as part of the High-Luminosity LHC (HL-LHC) project~\cite{hl-lhc}.
This upgrade will enable high-statistics physics searches, but it will also increase the event rate. In order to cope with the higher rate, the electronics will be upgraded and new trigger logic will be developed.
In the high-luminosity LHC-ATLAS experiment, large-scale FPGAs will be introduced to implement advanced and complex trigger logic.
Furthermore, with the recent adoption of High-Level Synthesis (HLS) and machine learning, the complexity and opacity of the logic and firmware have increased, necessitating rigorous validation.

Software-based validation methods, such as Vivado~\cite{vivado} simulation, are suitable for validating basic firmware behavior, however, their long execution times render them impractical for large-scale validation.
Similarly, hardware-based validation using actual devices or evaluation boards is limited by hardware availability and the complexity of implementing input/output interfaces.
A tool is needed trigger logic to validate the trigger logic on the FPGA.
Therefore, we have developed a high-speed and flexible trigger logic firmware validation system using FPGA accelerators.

In ATLAS~\cite{atlas2008}, the Virtex UltraScale+ FPGA~\cite{virtex_ultrascale} from AMD/Xilinx is employed for its high performance and reliability~\cite{atlasMuon2017}. To ensure architectural consistency, the validation system has been developed using AMD Alveo FPGA accelerator cards, which utilize the same Virtex UltraScale+ FPGA family. By employing hardware acceleration with Alveo FPGA accelerators, it becomes feasible to execute trigger logic directly on FPGAs.
The validation system exclusively employs Vivado and Vitis~\cite{vitis} -- the primary software tools used in firmware development for the ATLAS experiment -- allowing it to be integrated without modifications to the existing development environment.
Moreover, by avoiding the use of expensive logic boards, trigger logic development can be carried out in parallel at multiple research institutes.
Unlike existing accelerator-based validation systems such as FireSim~\cite{FireSim}, our system is designed for local installation without any changes to the current development environment.

\section{FPGA accelerator}
FPGA accelerators are extension cards that incorporate an FPGA and connect to the host PC’s motherboard via PCI-Express. These accelerators are equipped with DDR memory, which stores data transferred from the host and facilitates high-speed communication between the host and the FPGA. Data exchange between the host PC's DDR and the FPGA accelerator's DDR is performed via Direct Memory Access (DMA) over PCI-Express, while data transfer between the DDR and the FPGA is managed through the AXI interface~\cite{axi}. Once computations are executed by the implemented circuit, the processed data is transferred back to the CPU in the reverse order. The FPGA accelerator receives input parameters for firmware execution through user-defined control code written in C++.  

Table~\ref{fig:comparison} presents a comparative analysis of the specifications and costs of various FPGA accelerators alongside standalone FPGA chips.
The reason why the FPGA accelerator card is cheaper than a standalone FPGA chip lies in the broader market, such as data centers, and reduced manufacturing costs due to mass production.
On the other hand, standalone FPGA chips with high-end resources are only required for limited purposes, such as HEP experiments, and are expensive.
Actually, the FPGA core XCU250 used in the Alveo U250 has same amount of resources on the XCVU13P which is one of the used FPGA chips for ATLAS.
Although the proposed validation system is adaptable to any AMD FPGA accelerators, the Alveo U200 was selected as an example in this paper.

\begin{table}[htbp]
    \centering
    \caption{\label{fig:comparison} The performance and price comparison between Alveo FPGA accelerators and standalone FPGA chips. 
    % \textcolor{red}{The price of accelerators is more than ten times cheaper than that of general-purpose FPGA cores with same resource.}
    % Accelerators are priced low due to intense price competition and mass production. As a result, they are more than 10 times cheaper than FPGA chips of the same architecture.
    The prices of accelerators are more than 10 times cheaper than standalone FPGA chips with same resource.}
    The Alveo U200 was bought at the academic price on May 25, 2021, and the FPGA chip was quoted at the academic price estimated on January 2025. The price of the Alveo U250 refers to the current market rate as of January 2025~\cite{U250_value}.
    % The Alveo U200 was purchased at the academic price on May 25, 2021.
    \label{tab:hogehoge}
    \begin{threeparttable}
    \begin{tabular}{ccccc}
        \hline
        & \multicolumn{2}{c}{FPGA accelerator~\cite{Alveo_spec}} & \multicolumn{2}{c}{FPGA chip~\cite{UltraScale_spec}} \\
        & Aleo U200 & Alveo U250 & XCVU9P & XCVU13P\\ \hline
           LUTs (K) & 1,182 & 1,728 & 1,182 & 1,728 \\
           Registers (K) & 2,364 & 3,456 & 2,364 & 3,456 \\
           UltraRAM Blocks  & 960 & 1280  & 960 & 1280 \\
           Price (JPY) & ¥ 494,450 & ¥ 1,637,129  & ¥ 8,111,800 & ¥ 11,421,800\\ \hline
    \end{tabular}
    % \begin{tablenotes}
    % \item[a] Prices as of 25 May 2021.
    % \item[b] Prices as of 6 January 2025.
    % \end{tablenotes}
    \end{threeparttable}
\end{table}

% \begin{figure}[htbp]
% \centering % \begin{center}/\end{center} takes some additional vertical space
% \includegraphics[width=.6\textwidth,origin=c]{Alveo_U200_components}
% \caption{\label{fig:component} The communication format between the PC and the Alveo U200, and the components of the XCU200. The DMA controller implemented in the Static Region executes communication via PCI-express. \cite{e}}
% \end{figure}

\section{Overview of the developed trigger logic validation system.}

%\sout{The system development was conducted using the AMD/Xilinx Vivado and Vitis applications.}

\begin{figure}[htbp]
\centering % \begin{center}/\end{center} takes some additional vertical space
\includegraphics[width=.4\textwidth,origin=c]{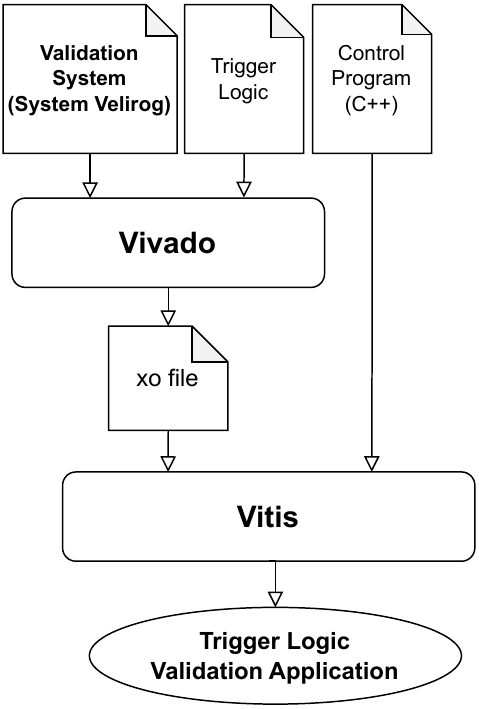}
\caption{\label{fig:flow} 
The validation system is implemented using System Verilog, one of the HDLs, and wraps the trigger logic to generate a single Vivado project. The project is packaged as a single IP and converted into a Xilinx Object (xo) file. In Vitis, the xo file is used for circuit synthesis to the FPGA, and combined with control C++ code to generate an application. The trigger logic validation is performed by inputting physical data generated through MC simulations into this application.}
\end{figure}

The system development was carried out using Vivado and Vitis applications, which are the same toolchain used for trigger logic development in the ATLAS experiment.
The trigger logic firmware, described in an HDL, was wrapped using the developed validation system to create the project.
Figure~\ref{fig:flow} illustrates the simplified steps involved in validating the trigger logic using the developed system.

%\sout{The validation system was implemented using System Verilog, one of the HDLs. The created project was packaged as an IP using Vivado and output as an \texttt{xo} file, one of the object files. An application was developed in C++ to control the input and output of data to and from the accelerator. This application was combined with the \texttt{xo} file and implemented as a single application using Vitis. Vitis implements the logic described in the \texttt{xo} file as a circuit within the FPGA. The circuit implementation time varies depending on the size of the trigger logic, ranging from a minimum of 2 hours to as long as 12 hours.}

\begin{figure}[htbp]
\centering % \begin{center}/\end{center} takes some additional vertical space
\includegraphics[width=.9\textwidth,origin=c]{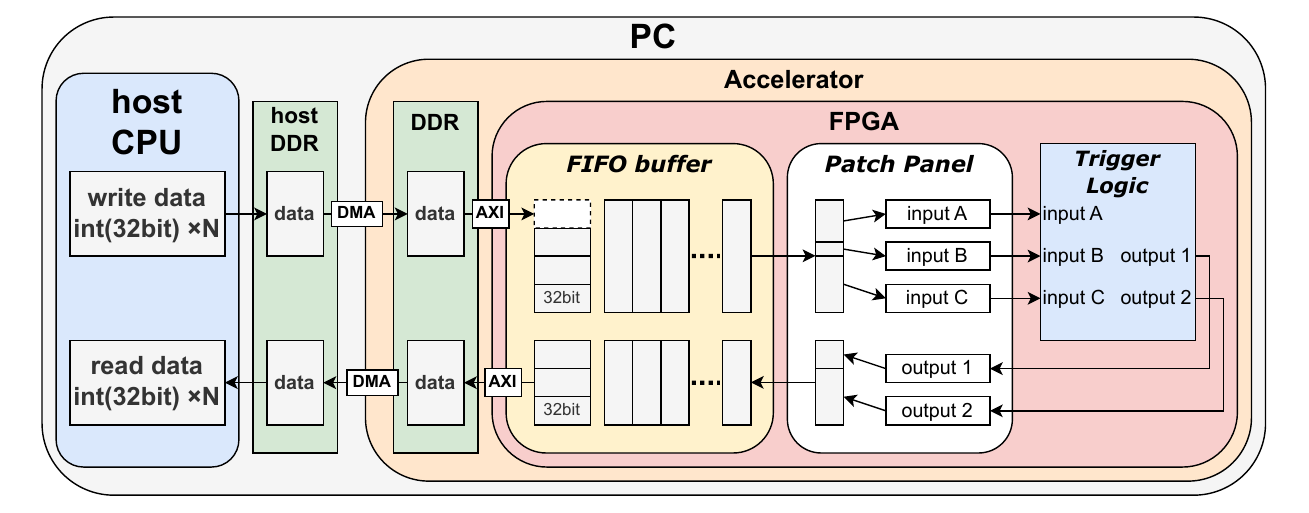}
\caption{\label{fig:diagram} Diagram of the developed validation system. The system manages communication between the FPGA and the host CPU, handles data merging and the distribution of input/output signals to the trigger logic for smooth validation.
Data generated by the host PC is transferred from the host PC to the DDR of the FPGA accelerator via DMA and then input to the FPGA through the AXI interface. All data communication is executed in 32-bit units. The FIFO buffer logic adjusts the timing and merges the data, preparing input signals of one event for the trigger logic. The merged data in the FIFO buffer logic is distributed to the trigger logic input signals by the Patch Panel logic. The output signals from the trigger logic are available on the host PC.}
\end{figure}

The trigger logic validation system we developed consists of several components. Figure~\ref{fig:diagram} shows the block diagram of the validation system, which includes a section that transfers data from the DDR within the accelerator to the FPGA via the AXI interface, as well as a FIFO buffer logic and patch panel logic that handle the storage and input/output to the trigger logic.
%\sout{The part that transfers data from the DDR within the accelerator to the FPGA via the AXI interface, }
%\sout{the FIFO buffer logic that combines and stores the transferred data, and the Patch Panel logic that aligns the combined data from the FIFO buffer with the input of the trigger logic.} 
The validation system monitors the state of the trigger logic and employs control flags to manage data transmission and reception.
%\sout{flag information from the FPGA to the DDR to control the AXI interface. In contrast, it manages data transmission and reception based on the flag information received from the DDR.}
%\sout{The FIFO buffer logic combines the input data, which is divided into 32-bit segments by the AXI interface, back into a single large data set. During output, it reverses this process, dividing the large data set into 32-bit segments and transferring them to the DDR. The Patch Panel logic distributes the large data set created by the FIFO buffer logic to match the input of the trigger logic. During output, it combines multiple outputs into one and sends them to the FIFO buffer.}
%\sout{Through the data control by the FIFO buffer logic and Patch Panel logic, inputs and outputs wider than 32 bits can be handled without issues for the trigger logic.}
Even if the bit width of the data transferred from the DDR differs from that of the data input to the trigger logic circuit, customized inputs can be provided through data latency adjustment by FIFO buffer logic and bit distribution by patch panel logic.
The process from data generation by the CPU, through computation by the trigger logic, to the return of the computation results to the CPU constitutes one cycle. 
In each cycle, the system transfers data with a size of up to $32\times4096$~ bits.
% From 1 to a maximum of 4096 clocks of data can be transferred simultaneously in each cycle.
% In such cases, 
The FIFO buffer logic aligns the data for one cycle after receiving whole data of one event, then sends to the Patch Panel logic.

%\sout{The validation system can repeatedly execute the input and output data by running this validation cycle multiple times. By treating the clock signal input to the trigger logic as one of the data inputs, the delay time from signal input to output can be measured.}
% The output results can be validated in text file or root file format.
% Additionally, they can be compared with the results of other software simulators to further validate the trigger logic.

\section{Implementation of trigger logic and measurement of validation system performance}
\begin{figure}[htbp]
\centering % \begin{center}/\end{center} takes some additional vertical space
\includegraphics[width=.9\textwidth,origin=c]{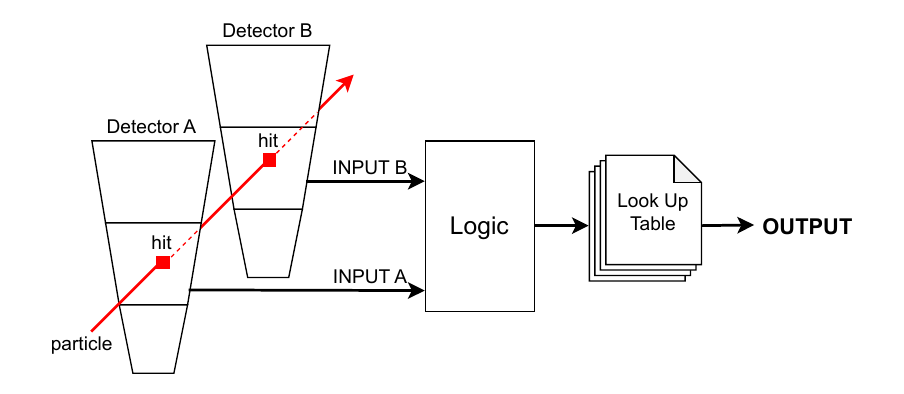}
\caption{\label{fig:example_logic} 
% Conceptual diagram of the trigger logic implemented in the validation system. This trigger logic is designed to extract pseudo-rapidity $\left( \eta \right)$ from the hit information of Detector A and Detector B, and obtain coincidences between the two detectors. The difference between $\eta_{A}$ and $\eta_{B}$ was calculated and entered into an LUT, with the result of the LUT serving as the final output.
Overview of the example logic implemented in the validation system. This logic processes hits passing through two detectors, converts them into a single address for the Look-Up Table, and then outputs the corresponding value.}

\end{figure}

As an example, we developed the trigger logic shown in Figure~\ref{fig:example_logic} using System Verilog and integrated it into the validation system. 
% \sout{
% In this validation, the value of $\eta_{A}$ was kept constant, while the value of $\eta_{B}$ was varied every four clock cycles.
% Upon executing the validation system with this trigger logic, Figure~\ref{fig:time} shows the time required for each part of one validation cycle.
% }

\begin{figure}[htbp]
\centering % \begin{center}/\end{center} takes some additional vertical space
\includegraphics[width=.6\textwidth,origin=c]{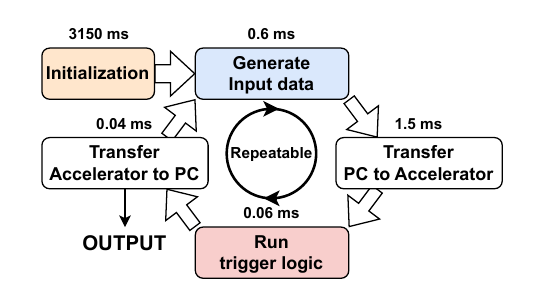}
\caption{\label{fig:time} The average time required for each part of the validation process. Initialization is only necessary at startup. One cycle of the validation loop corresponds to validating one event in the trigger logic.}
\end{figure}

%This trigger logic is intended for use in the HL-LHC ATLAS experiment. This logic implements a LUT to calculate the transverse momentum ($p_T$) from the difference in eta. Therefore, it is necessary to validate whether the LUT outputs the correct $p_T$) using MC data. We created $\eta_{A}$ and $\eta_{B}$, which are the hit coordinates of particles in the MC data, and input them into the trigger logic to measure the time required for validation.
To evaluate that the Look-Up Table in this logic is correctly implemented, we provided the trigger logic with hit data from detectors created by MC simulations and checked the output.
Figure~\ref{fig:time} shows the average time taken for data transfer and computation in the trigger logic. We evaluated the performance by preparing 25,000 bits of data per event.
The initialization for the device configuration takes approximately 3 seconds, and the validation cycle for 1,000 events takes 2.2 seconds

% In this validation, 50 clock cycles of data are input per cycle.
In a similar validation using Vivado simulation, due to the large scale of the trigger logic, it took 160 seconds from the start of the simulation to execution, and 58 seconds to process data of 1000 events. When comparing the data processing speed per event, ignoring the time for circuit synthesis, the validation system we developed achieves a significant speedup. The system demonstrates a speed improvement of approximately 1,300 times compared to Vivado simulations. Therefore, as demonstrated in this study, it becomes a very powerful tool when executing trigger logic using large-scale physical data generated by MC simulations.

%\sout{The output can be viewed as a text file in the format shown in Figure
%~\ref{fig:output}
%. Comparing the output results obtained in text file format with the results from Vivado simulation showed no discrepancies.}

% \begin{figure}[htbp]
% \centering % \begin{center}/\end{center} takes some additional vertical space
% \includegraphics[width=.8\textwidth,origin=c]{sample_logic}
% \caption{\label{fig:logic} Conceptual diagram of the trigger logic implemented in the validation system. This trigger logic is designed to extract pseudo-rapidity $\left( \eta \right)$ from the hit information of Detector A and Detector B, and obtain coincidences between the two detectors. The difference between $\eta_{A}$ and $\eta_{B}$ was calculated and entered into an LUT, with the result of the LUT serving as the final output.}
% \end{figure}

% \begin{figure}[htbp]
% \centering % \begin{center}/\end{center} takes some additional vertical space
% \hspace{1cm}
% \includegraphics[width=.7\textwidth,origin=c]{result_output}
% \caption{\label{fig:output} The output results obtained from the trigger logic are stored in text format using C++ code. In this figure, the output is captured every clock cycle. The latency of the logic used was 4 clock cycles, and the output results were obtained from a pre-implemented LUT.}
% \end{figure}
% \newpage
\section{Conclusion}
We have developed a flexible and high-speed trigger logic validation mechanism using FPGA accelerators. This validation mechanism has successfully optimized previously time-consuming and challenging validation processes, such as comprehensive LUT validation. The developed system seamlessly integrates data generated by MC simulations for input, facilitating the validation of trigger logic. The proposed FPGA accelerator-based validation mechanism achieved a speedup of approximately 26 times compared to Vivado simulations.

The validation mechanism can be executed on a single PC, which simplifies the setup process. This approach enables validation even without an actual FPGA board, greatly simplifying the setup of the validation environment. It also supports the direct use of HDL and HLS firmware code, enabling high-speed validation. Therefore, we believe that this mechanism can be applied across the field of particle physics experiments.

% \appendix
% \section{Some title}
% Please always give a title also for appendices.

% \acknowledgments

% This is the most common positions for acknowledgments. A macro is
% available to maintain the same layout and spelling of the heading.

% \paragraph{Note added.} This is also a good position for notes added
% after the paper has been written.


\acknowledgments
This work was partially supported by the Japanese Ministry of Education, Culture, Sports, Science and Technology, Grant-in-Aid for Scientific Research (16H06493 and 23H04511).

\end{document}